\documentclass{osa-article-preprint}

\journal{osajournal}

\usepackage[T1]{fontenc}
\usepackage{color}

\usepackage{soul} 

\articletype{Research Article}

\begin{document}

\title{How Dark is Dark? A Reflectance and Scattering Analysis of Black Materials}

\author{Ji\v r\'\i~Filip\authormark{*}, Radom\'\i r V\'avra}

\address{Czech Academy of Sciences, Institute of Information Theory and Automation, Praha, Czech Republic}

\email{\authormark{*}filipj@utia.cas.cz} 


\begin{abstract}
Black materials play a critical role in applications such as image registration, camera calibration, stray light suppression, and visual design. Although many such materials appear similarly dark under diffuse illumination, their reflectance behavior can differ substantially as a function of viewing and lighting geometry. Ultra-black materials achieve exceptional light attenuation but are often constrained by cost and mechanical fragility, motivating the evaluation of more robust and accessible alternatives. In this study, we employ a gonimetric measurement system to capture the isotropic bidirectional reflectance distribution function of a range of black materials, including the ultra-black reference Vantablack, commercially available alternatives such as Musou Black and black velvet, and standard matte black coatings. We analyze their reflectance characteristics in terms of diffuse and specular scattering, as well as total integrated scatter, to quantify angular-dependent reflection. In addition, we compare their perceptual appearance using physically based rendering driven by the measured BRDFs and a psychophysical evaluation of perceived darkness. Together, these analyses provide a comprehensive assessment of black materials that links reflectance properties to visual appearance and perceptual performance, enabling informed material selection for optical applications.
\end{abstract}


\section{Introduction}

Black materials are essential components in both consumer-grade optical systems and high-performance scientific instruments, where they serve to suppress stray light and minimize reflections \cite{Georgiev2019}. The required level of reflectance suppression varies by application, depending on factors such as signal-to-noise ratio constraints, device price, and tolerance to angular-dependent scattering. While basic black coatings may suffice for general-purpose use, more demanding systems -- such as photometric calibration targets, imaging sensors, or spaceborne optics -- require materials with extremely low directional reflectance to ensure accurate signal capture.

Angular dependence of reflectance is particularly critical in view- and light-direction sensitive applications, where unintended specular lobes or elevated off-specular scatter can impair registration accuracy or introduce imaging artifacts. To assess and compare such behaviors, Bidirectional Reflectance Distribution Function (BRDF) measurements provide a detailed characterization of material appearance across varying illumination and viewing geometries.

In this study, we conduct dense isotropic BRDF measurements of black materials spanning a range of types and applications -- from commercial matte paints and textiles to ultra-black surfaces such as vertically aligned carbon nanotube (VACNT) arrays. Measurements are acquired using a high-precision goniometric system in the visible (VIS) spectral range. We evaluate each material based on its diffuse and specular reflectance components, total integrated scatter (TIS), and rendered visual appearance using physically-based rendering. The results offer practical guidance for selecting black materials in optical and imaging systems with different performance and cost requirements.

\section{Related work}

The reflectance properties of black materials have long been studied due to their importance in applications such as optical calibration, stray-light suppression, remote sensing, and visual appearance control. Early work by Griner~\cite{Griner1979} and Newell and Keski-Kuha~\cite{Newell1997} established the importance of BRDF measurements for characterizing low-reflectivity coatings used in high-precision optical systems.

Subsequent work has expanded both the spectral range and angular resolution: Marshall et al.~\cite{Marshall2014} and Zeng et al.~\cite{Zeng2019} reported extensive measurements of total, diffuse, and specular reflectance for a variety of black coatings across ultraviolet, visible, and near-infrared wavelengths. Zeidler et al.~\cite{Zeidler2019} further investigated angular scattering distributions of highly absorptive surfaces, emphasizing the role of off-specular scattering in stray-light performance and highlighting the need for detailed angular characterization beyond hemispherical reflectance alone. 

Several studies have focused on detailed BRDF characterization and scattering behavior of specific material classes. Lu et al.~\cite{Lu1998} analyzed the bidirectional reflectance of black velvet, revealing its distinctive grazing-angle scattering behavior associated with fibrous surface microstructure. Related insights into structural mechanisms of optical darkness are provided by Spinner et al.~\cite{Spinner2013}, who showed that the extreme black appearance of the snake \emph{Bitis rhinoceros} results from hierarchical micro- and nanostructured surface features that enhance light trapping and suppress directional reflection. Shirsekar~\cite{Shirsekar2018} presented goniometric BRDF measurements of Aeroglaze Z302, examining the influence of coating thickness and incident angle on reflectance in the visible range. Strumik~\cite{Strumik2024} further developed a phenomenological BRDF model for Acktar Magic Black\textsuperscript{\textregistered}, providing insight into specular peak structure and off-specular energy redistribution.
Material durability and long-term stability have also been investigated, particularly in the context of space and remote sensing applications. Dury et al.~\cite{Dury2007} examined the aging behavior of common black coatings, including velvet-based materials, over a broad spectral range. Zeng et al.~\cite{Zeng2019} similarly reported environmental effects on optical performance, highlighting practical considerations that complement purely optical characterization.

Ultra-black materials based on vertically aligned carbon nanotubes (VACNT) have been reviewed extensively by Lehman et al.~\cite{Lehman2017}, who documented their exceptional absorptance and unique microstructure. While such materials represent an upper bound in achievable blackness, their extreme absorption poses challenges for BRDF measurement and practical deployment. Recent advances in flexible and more robust superblack materials have been reported by Yang et al.~\cite{Yang2025}, demonstrating continued interest in alternatives to fragile VACNT-based coatings.

Overview by Surface Optics Corporation~\cite{SurfaceOptics2024}, summarize current best practices in BRDF and BSDF measurement, emphasizing the importance of precise angular sampling, spectral control, and normalization underpining the goniometric acquisition strategy adopted in this paper. In contrast to prior studies that primarily focus on physical reflectance metrics, our work combines dense BRDF measurements in the visible spectrum with total integrated scatter (TIS) analysis, physically based rendering, and psychophysical evaluation. This integration establishes a direct link between angular reflectance behavior, its visual appearance, and perceived darkness, enabling practical guidance for selecting black materials across a range of optical applications.

\section{Materials and methods}

\subsection{Test materials}
\begin{figure}[!h]
\begin{center}
\renewcommand{\arraystretch}{0.4} 
\begin{tabular}{p{1.8cm}p{1.8cm}p{1.8cm}p{1.8cm}p{1.8cm}p{2.0cm}}
\small{Vantablack} & \small{Musou paint} & \small{black velvet} & \small{Musou fabric} & \small{acryl paint} & \small{chalkb. paint}\\
\includegraphics[width=0.15\columnwidth]{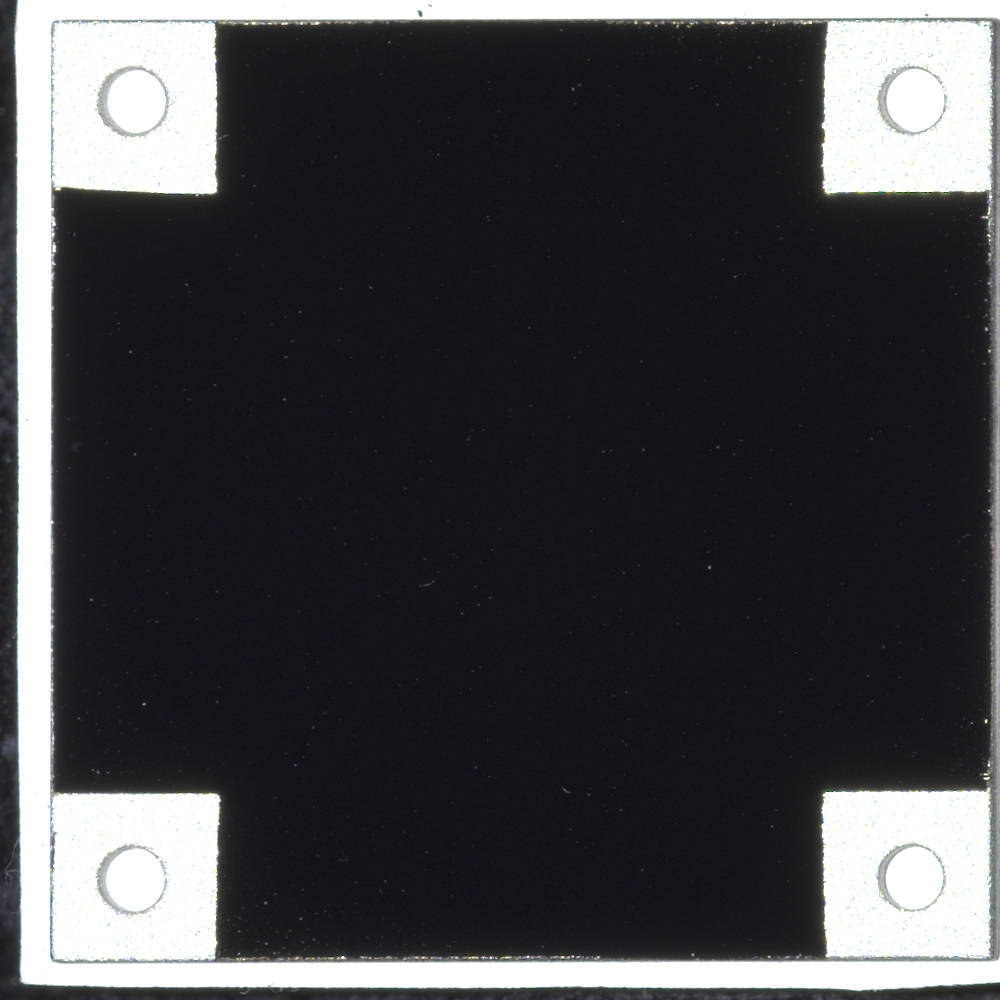} &
\includegraphics[width=0.15\columnwidth]{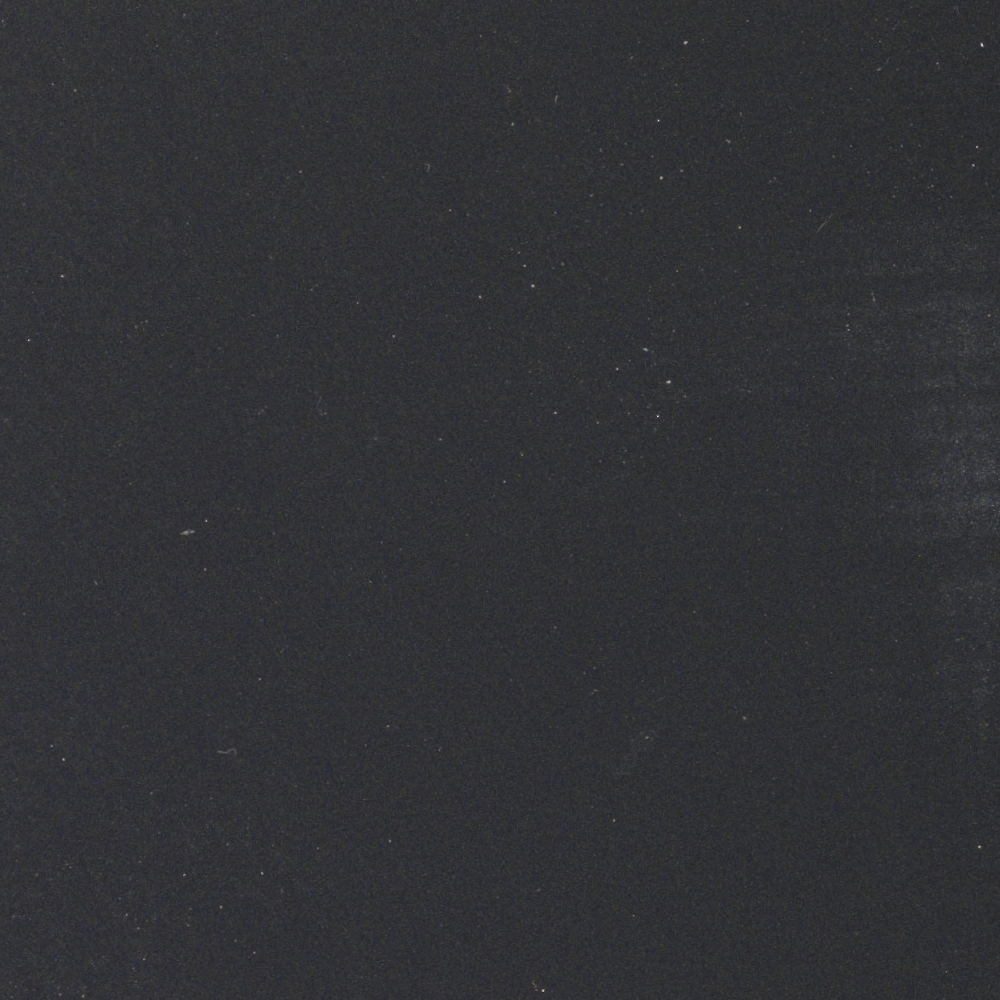} &
\includegraphics[width=0.15\columnwidth]{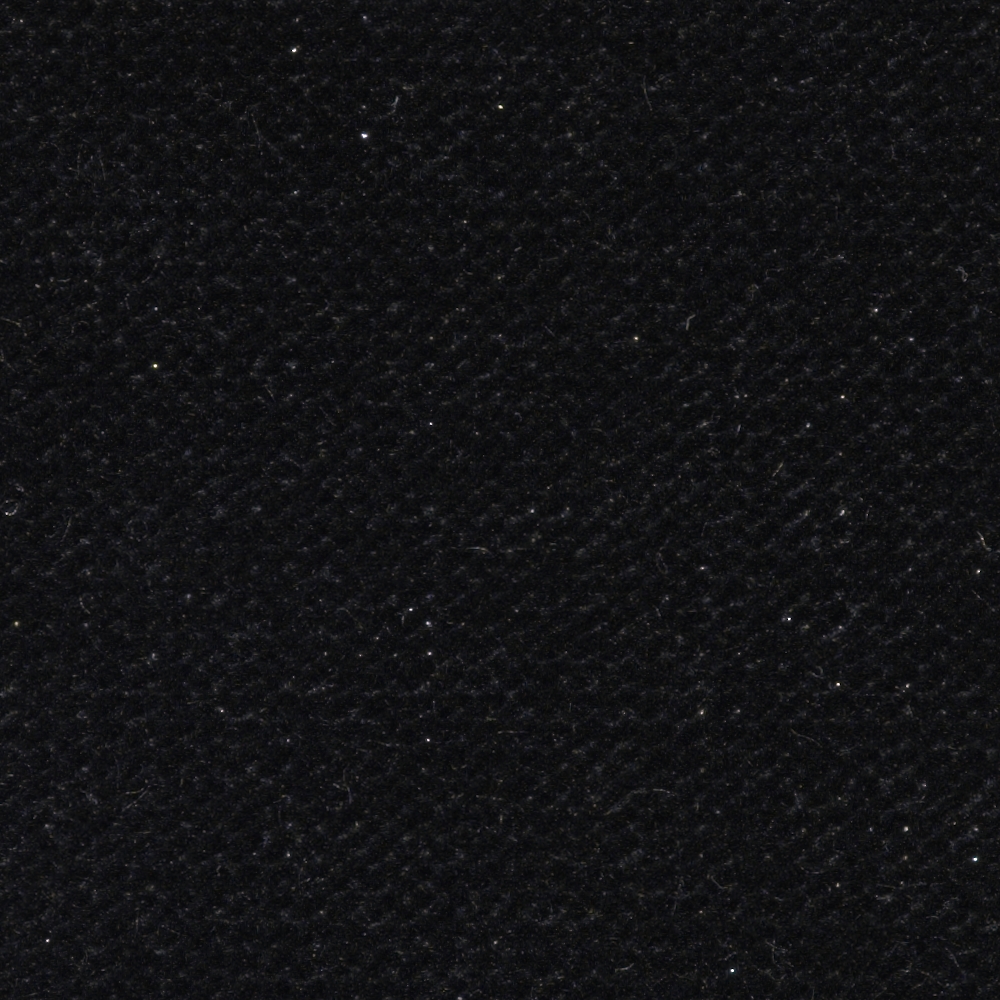} &
\includegraphics[width=0.15\columnwidth]{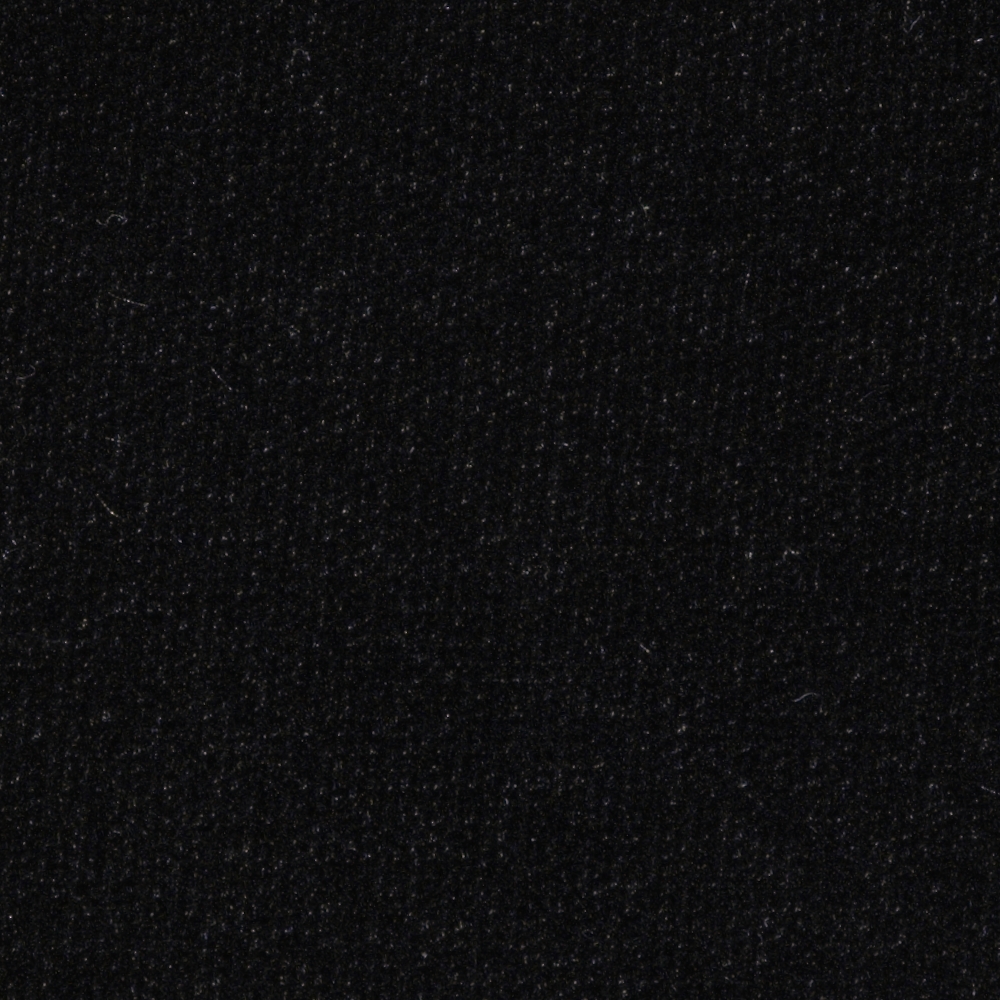} &
\includegraphics[width=0.15\columnwidth]{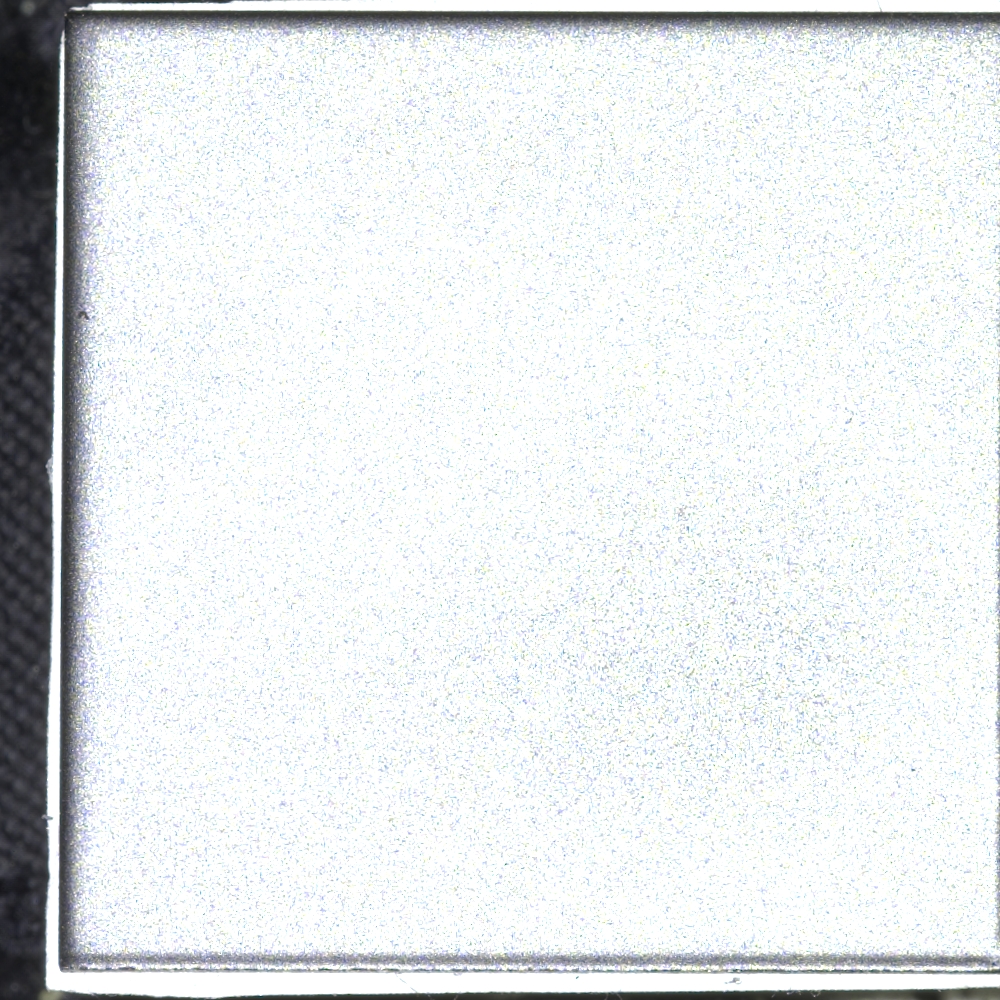} &
\includegraphics[width=0.15\columnwidth]{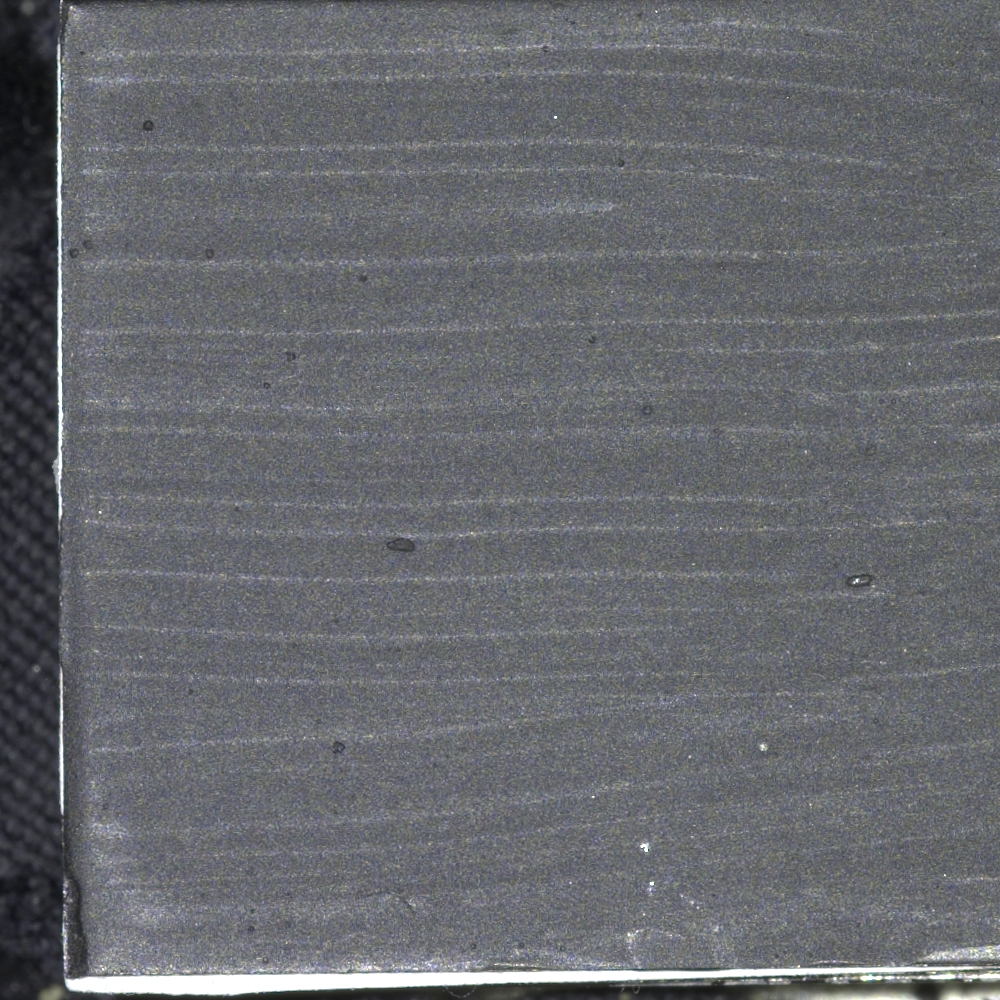} \\
\end{tabular}
\end{center}
\vspace{-5mm}
\caption{\label{fig:photo}Tested samples (40$\times$40 mm) at specular configuration $\theta_i/\theta_v$ = 45$^\circ$/45$^\circ$.}
\vspace{-2mm}
\end{figure}

In this paper we compared properties of material  featuring vertically aligned carbon nanotubes with three coatings and two fabric materials as shown in Fig.~\ref{fig:photo}:

\emph{Vantablack} -- One of the darkest materials known, Vantablack \cite{VantablackDatasheet} consists of VACNT \cite{Lehman2017} absorbing over 99.965\% of visible light, appearing as an almost featureless void. It is applied via specialized vacuum deposition and not available for consumer use.

\emph{Musou paint (brush-on)} -- a commercial ultra-black acrylic paint \cite{MusouSpecs2024} that absorbs up to 99.4\% of visible light when applied by brush, creating a dramatically matte surface with minimal specular reflection. 

\emph{Acrylic paint (spray)} -- a widely available matte paint with a low-gloss finish that provides moderate blackness and diffuse reflection in aesthetic/utility applications.

\emph{Chalkboard paint (brush-on)} -- a brush-applied black coating, offering a slightly glossy, durable surface that reflects more light than dedicated or ultra-black paints. 

\emph{Musou fabric} -- a deep-black synthetic fabric \cite{MusouSpecs2024} with a non-reflective pile surface designed for optical shielding and stray light suppression, offering near-total light absorption across a broad range of angles.

\emph{Black Velvet} -- a soft fabric with deep, angle-dependent blackness due to its light-trapping fibers; commonly used in optical benches and theatrical settings to reduce reflections.

\subsection{Measurement setup}

BRDF measurements were acquired using the UTIA goniometric system~\cite{filip13brdf}, which enables dense sampling of illumination and viewing directions with high angular precision. The setup consists of a stationary sample mounted on a precision rotation stage and two independently actuated arms carrying the illumination source and the imaging device. The illumination arm provides two rotational degrees of freedom, while the camera arm provides one, allowing acquisition of a wide range of illumination--viewing geometries. The verified angular positioning repeatability across all axes is 0.03$^\circ$. Within the paper, we represent illumination and viewing directions ($\omega_i, \omega_v$) as vectors in spherical coordinates relative to the surface normal. Elevation angles are denoted as $\theta_i$, $\theta_v$, and relative azimuthal angle $\varphi= |\varphi_i - \varphi_v|$. 

Unpolarized illumination is provided by a high-power LED source (Cree XML with a nominal luminous flux of 280 lm at 3.0 A) equipped with dedicated optics to produce a narrow and spatially uniform beam. High dynamic range acquisition is achieved by combining adaptive exposure times with adjustable illumination intensity controlled via the LED drive current. The light is positioned 1 m from the sample.

Images are captured using an industrial full-frame RGB camera (AVT Pike 1600C) with a 14-bit CCD sensor and a native resolution of $4872 \times 3248$ pixels. The camera is positioned approximately 2 m from the sample and equipped with a 180 mm focal length lens, yielding a spatial resolution of up to approximately 650 dpi at the sample plane. Owing to the low reflectance of the samples, the maximum exposure time was 2.6 s. The camera's spectral range is 380-700 nm with RGB channels peaks at 480, 520, and 615 nm.

System calibration includes geometric alignment, radiometric normalization, and colorimetric calibration. Sensor defects are detected and corrected in the raw image data. Optical vignetting is estimated from images of a uniform white reference and compensated accordingly. Colorimetric calibration is performed using a $3 \times 3$ linear transformation derived from measurements of an X-Rite color target in the CIE XYZ color space. All images are processed and stored in floating-point HDR format. 
BRDF values are obtained by averaging over a 4$\times$4 mm central image region and reported in relative units.
Images acquired from different viewing directions are registered at the pixel level using fiducial markers surrounding the measured sample. 

\subsection{Angular sampling}

To obtain sufficient sensitivity to reflectance changes we selected relatively dense sampling of azimuthal angles while keeping modest sampling step in lighting polar angle.
Each sample is measured for following geometries: Illumination elevation angles $\theta_i=$0$^\circ$, 15$^\circ$, 30$^\circ$, 45$^\circ$, 60$^\circ$, 75$^\circ$, 80$^\circ$, 85$^\circ$. Azimuth of all illumination directions is set to 0$^\circ$. For each illumination direction the viewing directions on hemisphere are sampled in elevation angles $\theta_v$ in a range 0-85$^\circ$ with 1$^\circ$ step and azimuthal angles in a range 0-360$^\circ$ with 10$^\circ$ step. This configuration provides sampling using angular BRDF slice of 8 $\times$ 86 $\times$ 36 = 24768 samples per material. Note that in our angularly uniform sampling are all azimuthal samples for $\theta_v=0^\circ$ identical, so the effective number of samples is 24488 (acquisition time ~96 hours).

\subsection{Measures of material darkness}

A variety of metrics have been proposed to quantify the optical performance of dark materials. Owing to the dense angular sampling of our BRDF measurements, several complementary measures of material darkness can be evaluated in a consistent framework.

A simple descriptive metric is the relative luminance, summarized using selected percentiles (e.g., $P_{1}$, $P_{50}$, $P_{99}$) over all measured combinations of illumination and viewing directions. While this measure provides an overall indication of reflected intensity, it does not capture angular dependence and is therefore insufficient on its own for characterizing dark materials under directional illumination.

A more physically grounded quantity is the total hemispherical reflectance (THR), obtained by integrating the BRDF over the outgoing hemisphere for a given incident direction $\omega_i$:
\begin{equation}
\rho(\omega_i) = \int_{\Omega_o} f_r(\omega_i, \omega_o)\,\cos\theta_v\,\mathrm{d}\omega_v ,
\label{eq:thr}
\end{equation}
where $f_r$ denotes the bidirectional reflectance distribution function (BRDF), $\omega_v$ is the outgoing direction, and $\theta_v$ is the angle between $\omega_v$ and the surface normal. By definition, $\rho(\omega_i)=0$ corresponds to a perfectly absorbing surface, while $\rho(\omega_i)=1$ corresponds to an ideal perfectly reflecting diffuser. In practice, very dark materials typically exhibit $\rho(\omega_i) < 0.02$ over most illumination directions.

To obtain a single material-level summary measure, we compute an effective albedo by integrating the cosine-weighted hemispherical reflectance over the sampled illumination directions,
\begin{equation}
A = \int_{\Omega_i} \rho(\omega_i)\,\cos\theta_i\,\mathrm{d}\omega_i ,
\label{eq:albedo}
\end{equation}
where $\theta_i$ is the illumination polar angle. In our case, this integral is approximated numerically using the discrete set of measured illumination directions.

To characterize stray-light behavior, we evaluate the total integrated scatter (TIS)~\cite{harvey12total}, which quantifies the fraction of reflected energy scattered away from the specular direction. In practical TIS measurements, a specular exclusion cone with half-angle between approximately $2^\circ$ and $5^\circ$ is commonly employed for real-world black materials (we used $5^\circ$). We decompose the total reflected energy $R_t$ into diffuse ($R_d$) and specular ($R_s$) components and define
\begin{equation}
\mathrm{TIS} = \frac{R_d}{R_t} = \frac{R_d}{R_s + R_d}.
\label{eq:tis}
\end{equation}

Finally, the dense BRDF data enable direct analysis of near-specular behavior, including the angular width and shape of specular lobes. These characteristics provide additional insight into grazing-angle reflectance and off-specular scattering mechanisms that are not captured by hemispherical metrics alone.

\subsection{Rendering setup and reference material}

The captured BRDF data have sufficient dynamic range and angular resolution to enable photorealistic rendering of material appearance on arbitrary 3D objects. Such visualizations allow assessment of material behavior in practical scenarios, for example as coatings on optical components. To this end, we designed a simple test scene consisting of four spheres illuminated by a single point light source. The spheres are positioned along a circular arc, with the light source located at the center of the arc, ensuring a wide range of illumination and viewing directions.

\begin{figure}[!ht]
\begin{center}
\begin{tabular}{p{0.2cm}p{5.2cm}p{0.2cm}p{5.2cm}}
(a)& \includegraphics[width=0.4\columnwidth]{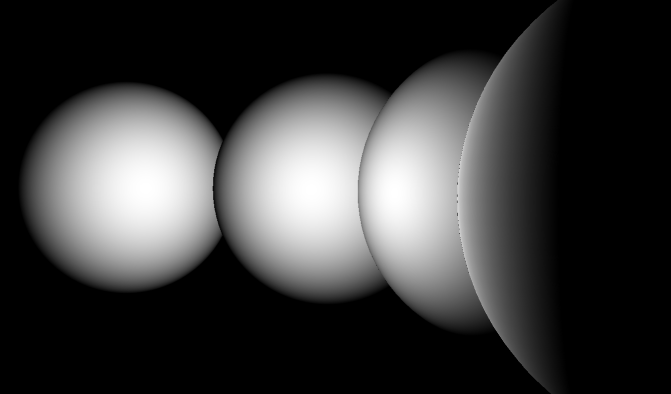} &
(b)& \includegraphics[width=0.4\columnwidth]{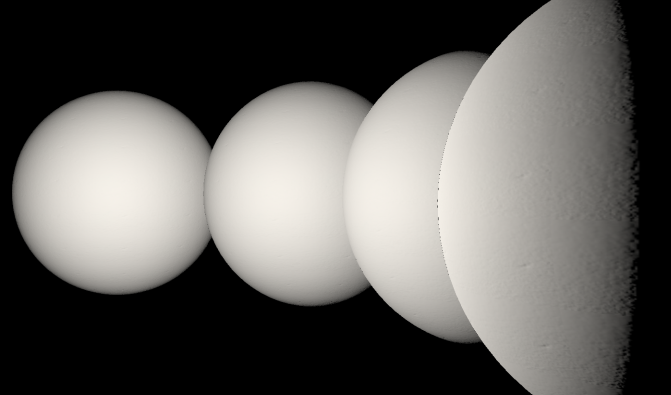} \\
\end{tabular}
\end{center}
\vspace{-6mm}
\caption{\label{fig:scene}Evaluated scene shown for (a) ideal Lamberian reflectance and (b) captured BTF of Spectralon data for nominal light intensity.}
\vspace{-3mm}
\end{figure}

Figure~\ref{fig:scene}(a) shows the scene rendered using an ideal Lambertian reflectance model. Figure~\ref{fig:scene}(b) shows a rendering of the captured Spectralon material \cite{LabsphereSpectralon}, which exhibits a hemispherical reflectance of approximately 99\% across the 250--2500 nm wavelength range under nominal illumination. While Spectralon is often treated as a near-Lambertian reference, deviations from an ideal cosine-weighted response are consistent with subsurface multiple scattering in the sintered PTFE structure. Owing to its high and well-characterized reflectance, Spectralon is used as an intensity reference when comparing low-reflectance materials.

\section{Results}

\subsection{BRDF maps and angular profiles}

\begin{figure}[!h]
\begin{center}
\renewcommand{\arraystretch}{0.4} 
\begin{tabular}{p{6.0cm}p{6.0cm}}
\small{Vantablack} & \small{Musou paint}\\
\includegraphics[width=0.46\columnwidth]{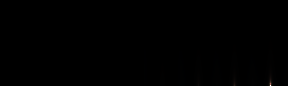} &
\includegraphics[width=0.46\columnwidth]{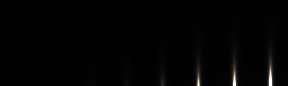} \\
\small{black velvet} & \small{Musou fabric}\\
\includegraphics[width=0.46\columnwidth]{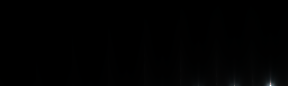} &
\includegraphics[width=0.46\columnwidth]{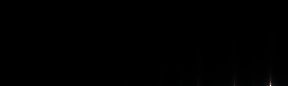} \\
\small{acryl paint} & \small{chalkboard paint}\\
\includegraphics[width=0.46\columnwidth]{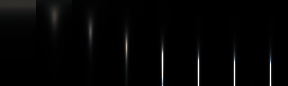} &
\includegraphics[width=0.46\columnwidth]{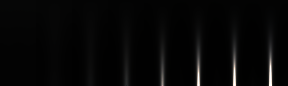} \\
\end{tabular}
\end{center}
\vspace{-5mm}
\caption{\label{fig:brdf}Captured BRDF slices visualizing 24 488 measured samples. Rows correspond to 36 discrete viewing polar angles, while columns represent azimuthal viewing angles for successive illumination polar angles $0^\circ, 15^\circ, 30^\circ, 45^\circ, 60^\circ, 75^\circ, 80^\circ$, and $85^\circ$.}
\vspace{-3mm}
\end{figure}

Captured BRDF slices in Fig.~\ref{fig:brdf} reveal pronounced differences in angular scattering behavior among the tested materials. The ultra-black reference \emph{Vantablack}, together with the fabric-based samples \emph{black velvet} and \emph{Musou fabric}, exhibits a broadly distributed reflected energy with no clearly discernible specular peak across the sampled illumination angles. 
 In contrast, the coating-based samples show significantly stronger and more localized specular responses that persist across multiple illumination polar slices (visible as consecutive bright vertical features). Among these, \emph{acrylic paint} produces the sharpest and most intense highlights, consistent with a pronounced specular component. \emph{Chalkboard paint} exhibits weaker but still clearly structured highlights with a broader angular spread, suggesting increased surface roughness. Notably, \emph{Musou paint} shows a markedly reduced specular response, with noticeable highlights appearing primarily at high illumination polar angles (near grazing incidence), indicating a more effective attenuation of near-normal specular reflection compared to other coatings.

\begin{figure}[!h]
\begin{center}
\includegraphics[height=0.32\columnwidth]{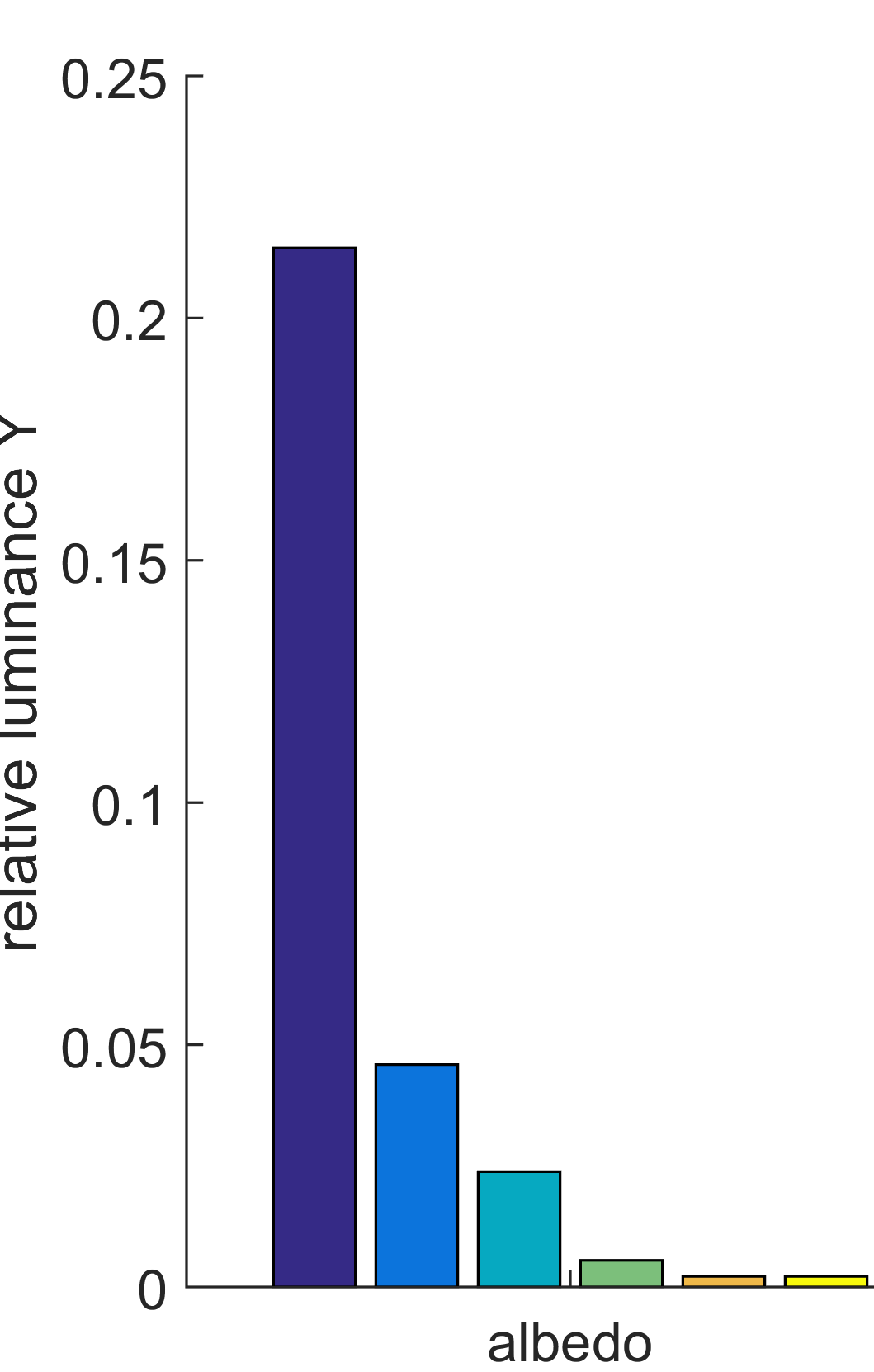} 
\includegraphics[height=0.32\columnwidth]{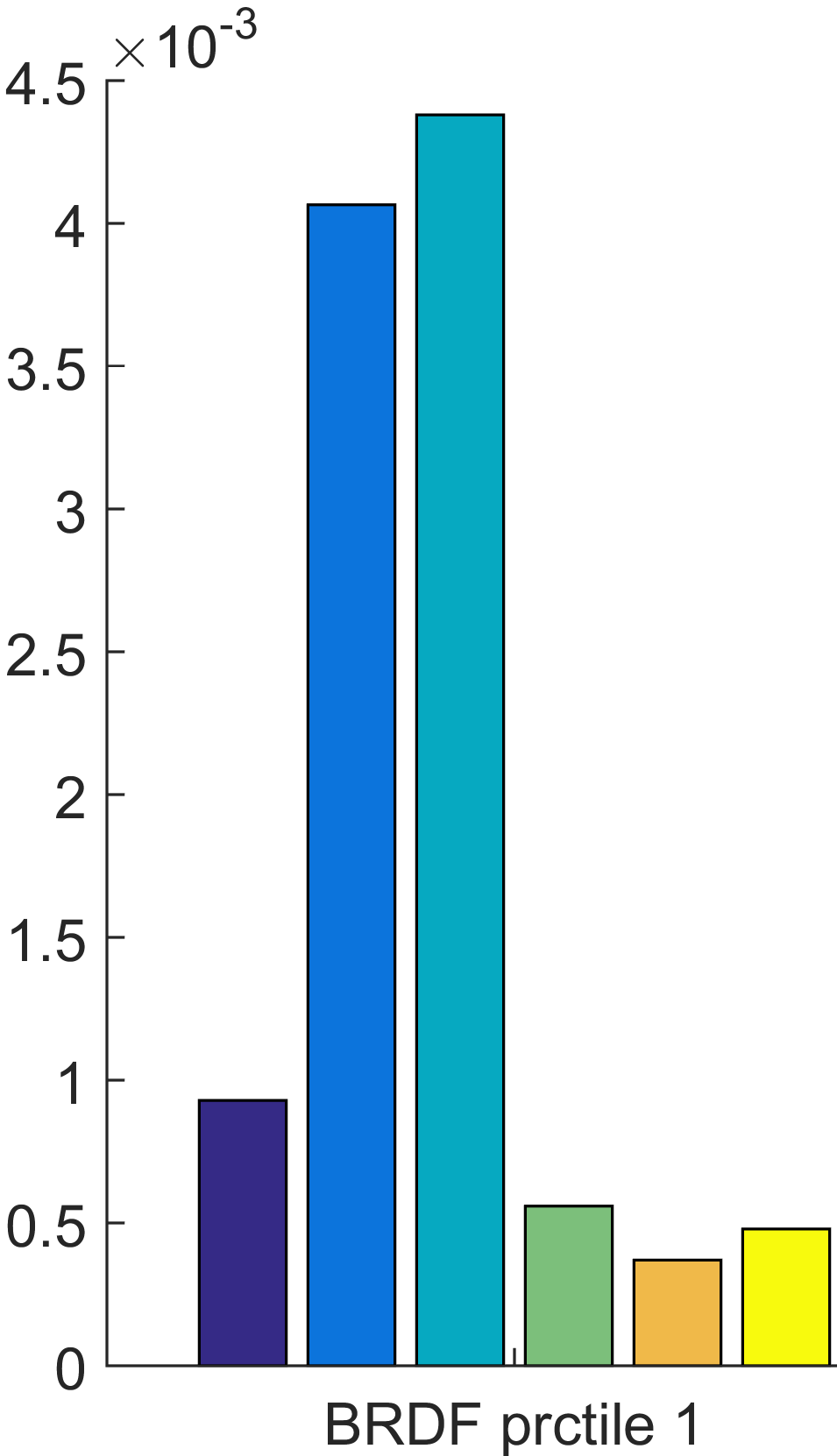} 
\includegraphics[height=0.32\columnwidth]{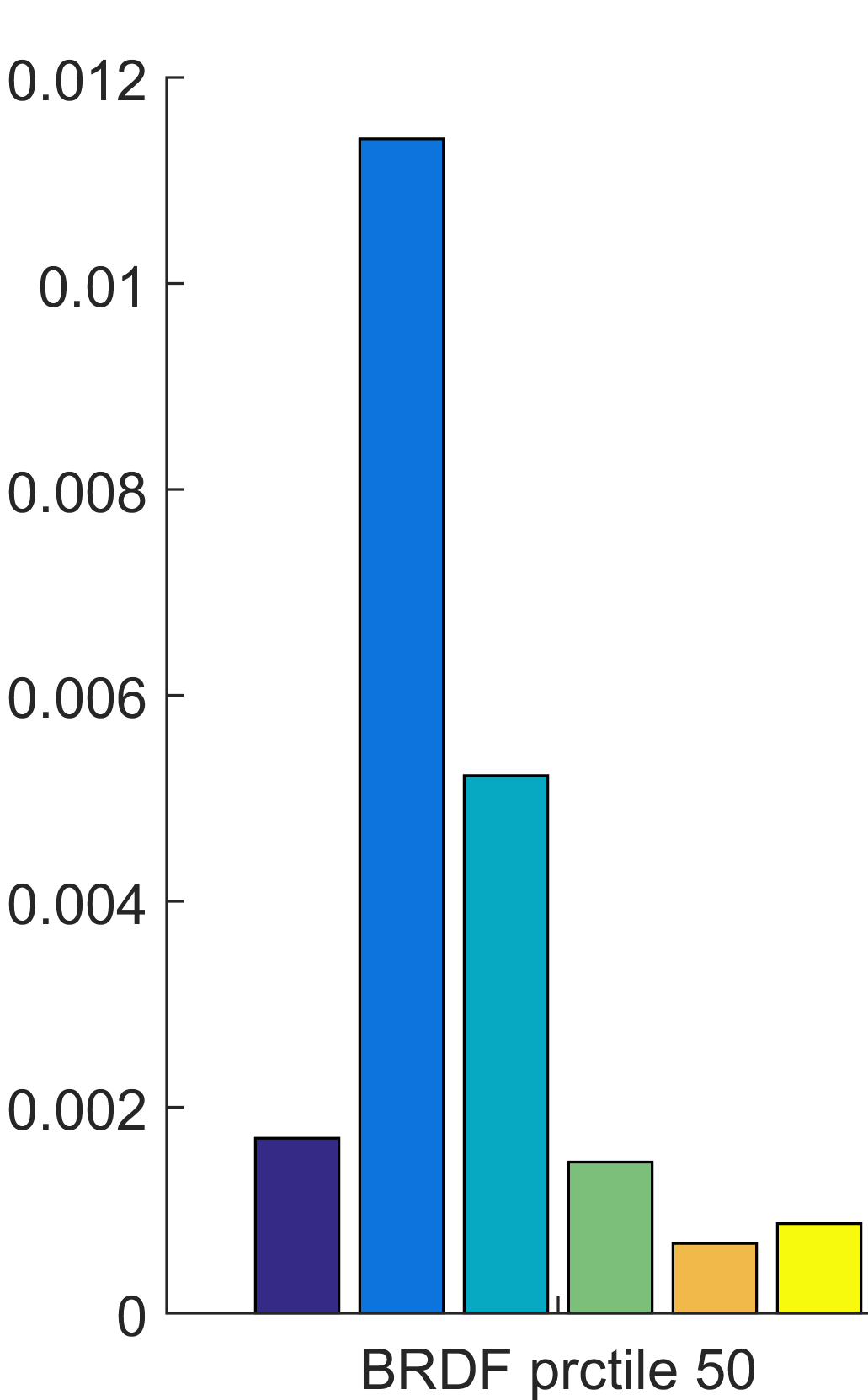} 
\includegraphics[height=0.32\columnwidth]{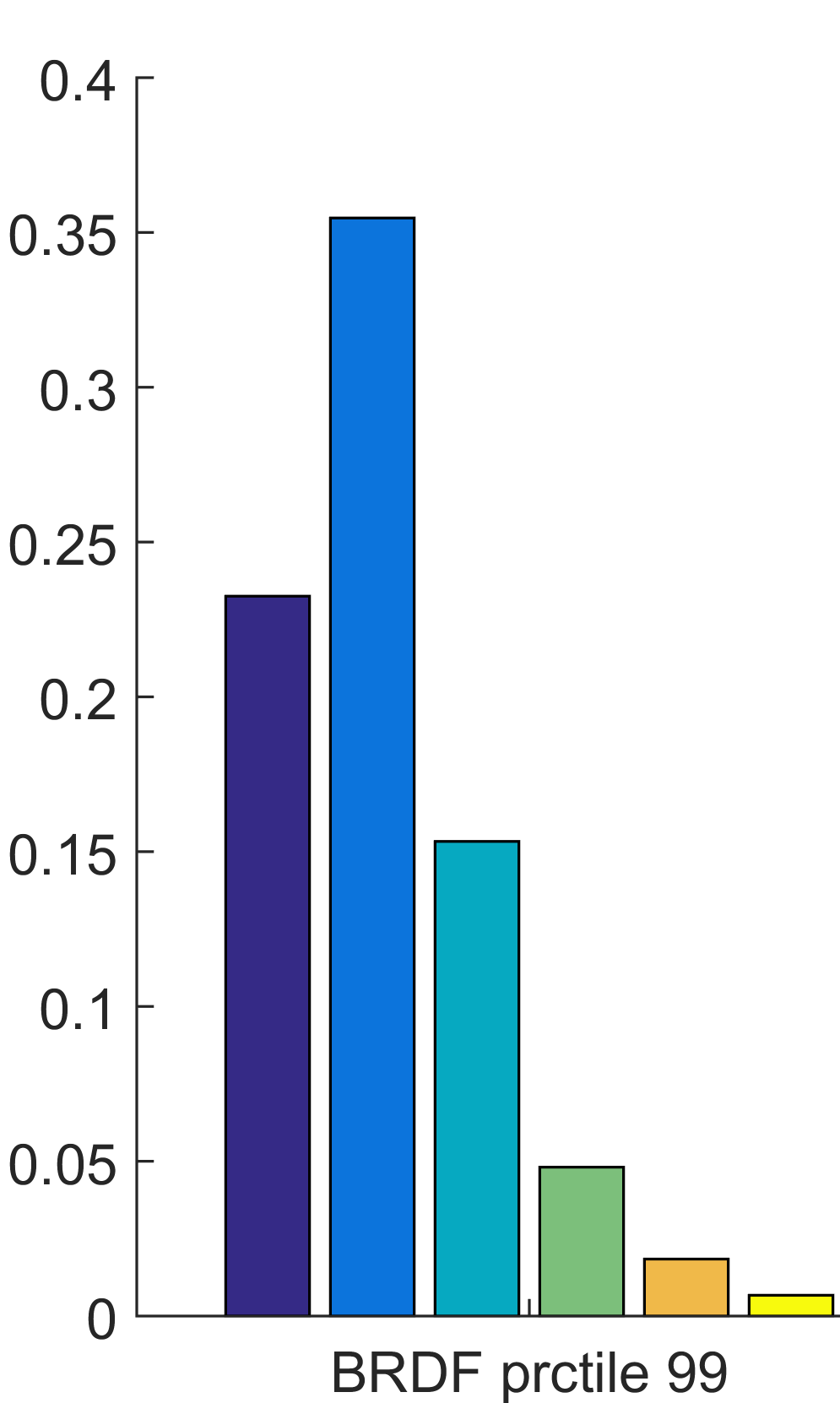} 
\includegraphics[width=0.16\columnwidth]{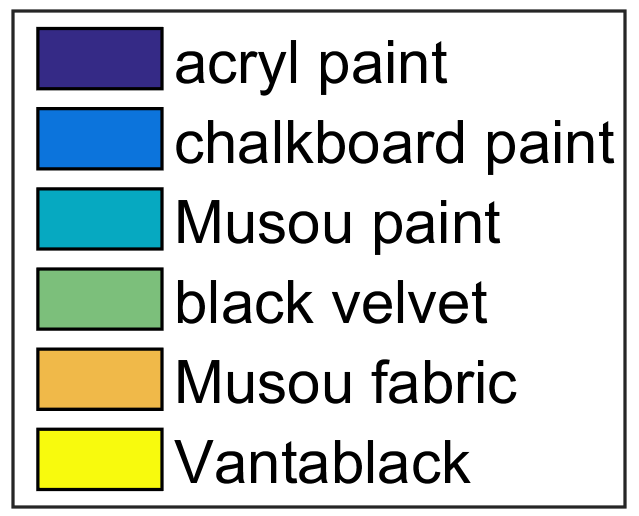} 
\end{center}
\vspace{-6mm}
\caption{\label{fig:lum}Albedo and percentiles of relative BRDF luminance.}
\vspace{-2mm}
\end{figure}

Fig.~\ref{fig:lum} summarizes the measured albedo values together with relative luminance statistics computed over the captured area. The luminance was obtained by weighted averaging of captured RGB values. The lowest albedo values are observed for \emph{Vantablack} and \emph{Musou fabric}, with \emph{black velvet} exhibiting a similarly low but slightly higher reflectance. In contrast, the coating-based samples show albedo values that are higher by approximately one order of magnitude, with \emph{acrylic paint} exhibiting the highest reflectance among all tested materials. A comparable trend is observed in the luminance statistics. To obtain robust measures and reduce sensitivity to isolated outliers, luminance values are summarized using the 1st (dark values), 50th (median), and 99th (bright values) percentiles. Notably, \emph{Musou fabric} achieves the lowest luminance values for both the 1st and 50th percentiles, indicating very effective suppression of diffuse and near-diffuse reflection. However, the 99th percentile values reveal that \emph{Vantablack} remains unmatched in attenuating high-intensity reflections, highlighting its superior performance in suppressing residual specular and off-specular scattering.

\begin{figure}[!ht]
\begin{center}
\begin{tabular}{p{3.5cm}p{3.3cm}p{3.3cm}p{3cm}}
\small{$\theta_i/\theta_v$ = 60$^\circ$/60$^\circ$}&
\small{$\theta_i/\theta_v$ = 75$^\circ$/75$^\circ$}&
\small{$\theta_i/\theta_v$ = 85$^\circ$/85$^\circ$}& \\
\includegraphics[width=0.3\columnwidth]{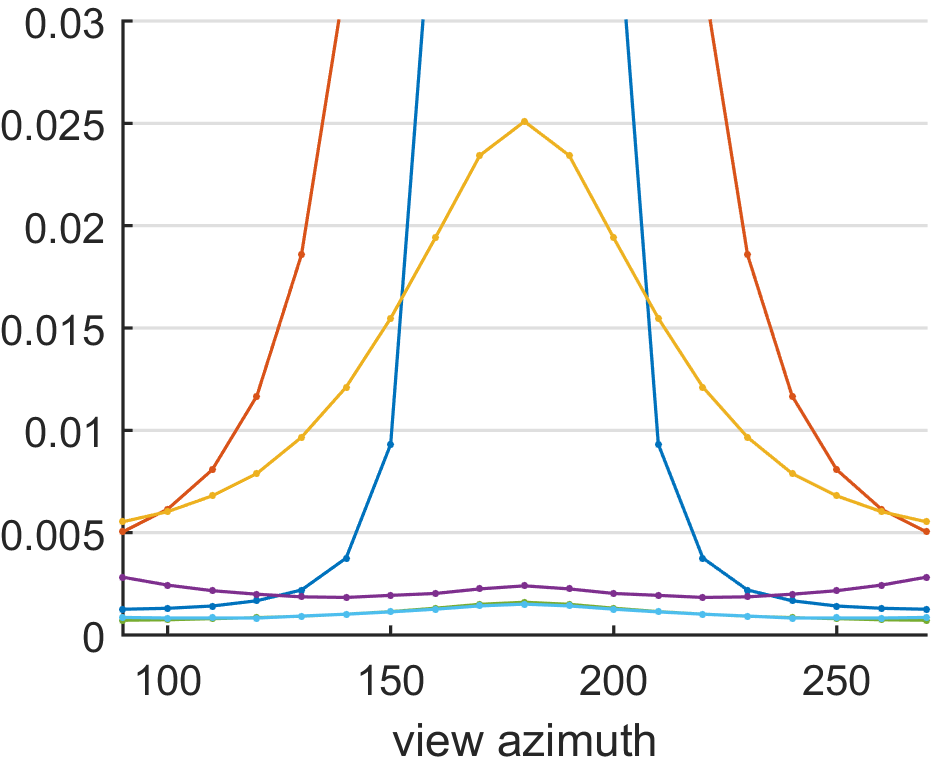} &
\includegraphics[width=0.3\columnwidth]{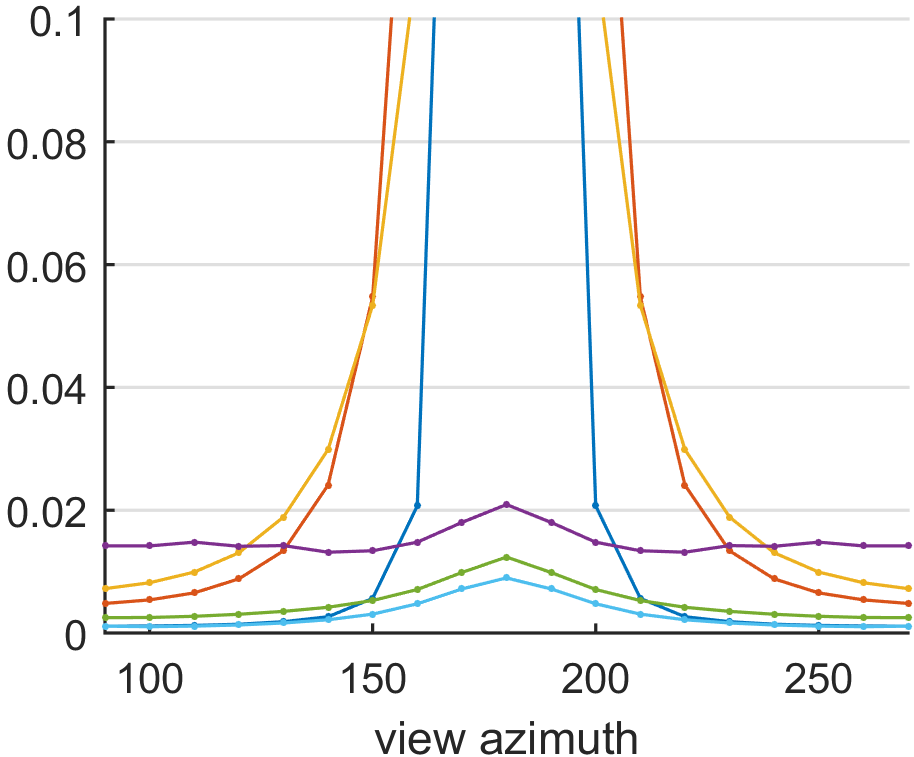} &
\includegraphics[width=0.3\columnwidth]{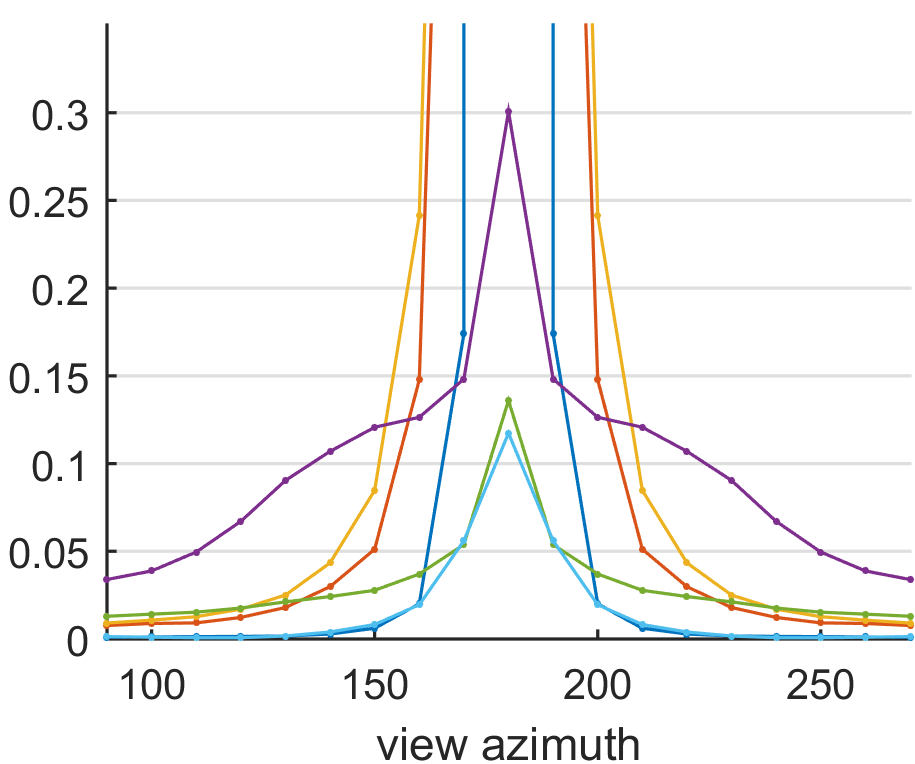} &
\includegraphics[width=0.16\columnwidth]{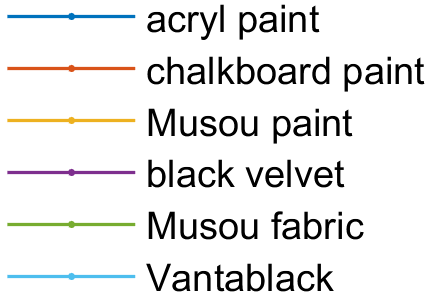} \\
\end{tabular}
\end{center}
\vspace{-6mm}
\caption{\label{fig:brdfHL}Profiles of specular reflection in BRDF for varying incident angles.}
\vspace{-3mm}
\end{figure}

Thanks to the dense angular sampling of the BRDF measurements, we can examine the detailed shape of specular scattering lobes by analyzing the relative luminance profiles for fixed and equal illumination and viewing polar angles $\theta_i = \theta_v$, plotted as a function of the viewing azimuth $\varphi_v$ in $10^\circ$ increments. Fig.~\ref{fig:brdfHL} shows representative azimuthal BRDF slices for increasingly grazing configurations. Across all configurations, \emph{Vantablack} and \emph{Musou fabric} consistently exhibit the lowest BRDF values, with weak and narrow specular responses, indicating strong suppression of directional reflection. \emph{Black velvet} follows closely, but with a noticeably higher diffuse baseline and broader angular spread, particularly at larger grazing angles.

In contrast, all coating-based samples show substantially stronger and wider specular lobes, whose peak intensity and angular extent increase with grazing illumination. Among these, \emph{acrylic paint} exhibits particularly broad lobes, maintaining relatively elevated BRDF values even at azimuthal angles more than $20^\circ$ away from the specular direction. \emph{Musou paint} displays a more controlled response compared to other coatings, with lower peak intensity and reduced angular spread, although its specular contribution remains markedly higher than that of the fabric-based and ultra-black materials.

\subsection{Total scattering analysis}
\begin{figure}[!ht]
\begin{center}
\begin{tabular}{p{3.9cm}p{3.9cm}p{3.9cm}}
\small{TIS} & \small{THR} & \small{$R_s$} \\
\includegraphics[width=0.32\columnwidth]{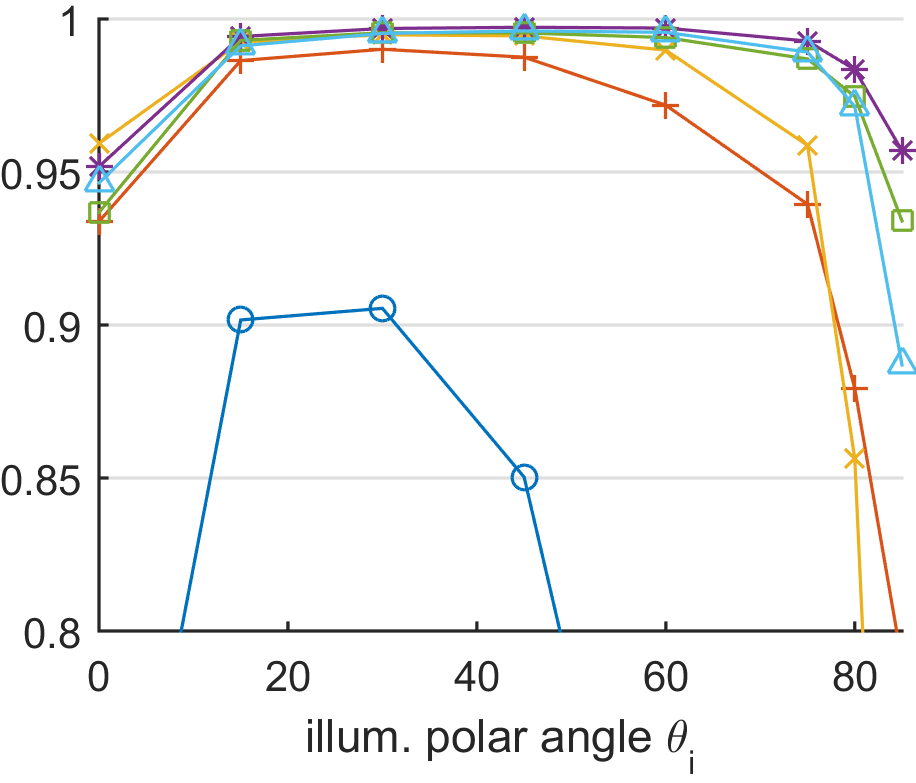} &
\includegraphics[width=0.32\columnwidth]{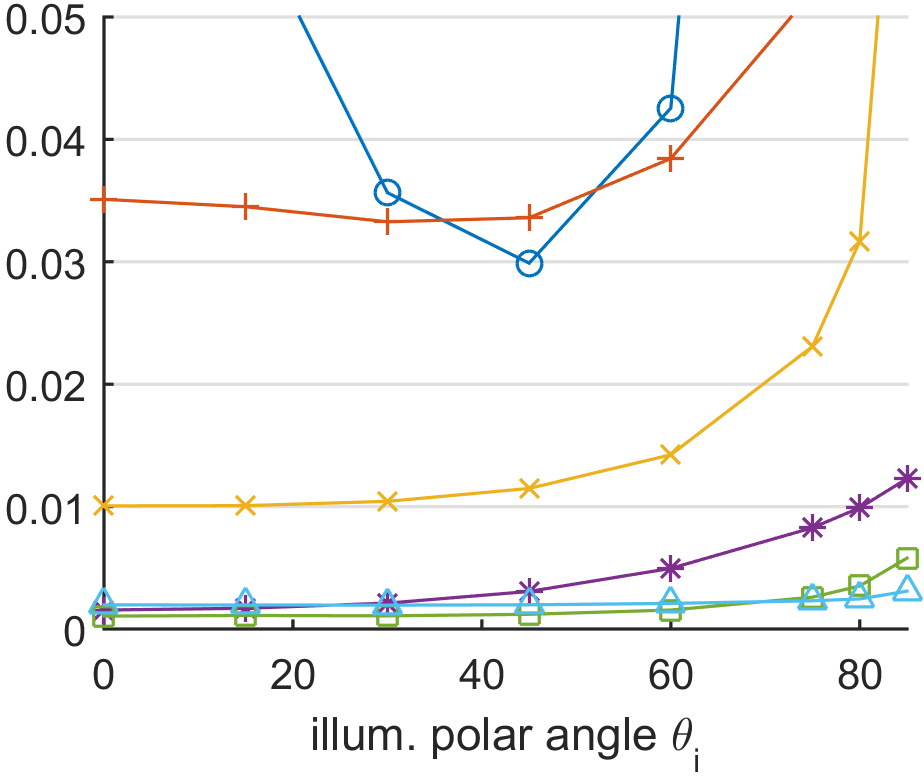}& 
\includegraphics[width=0.32\columnwidth]{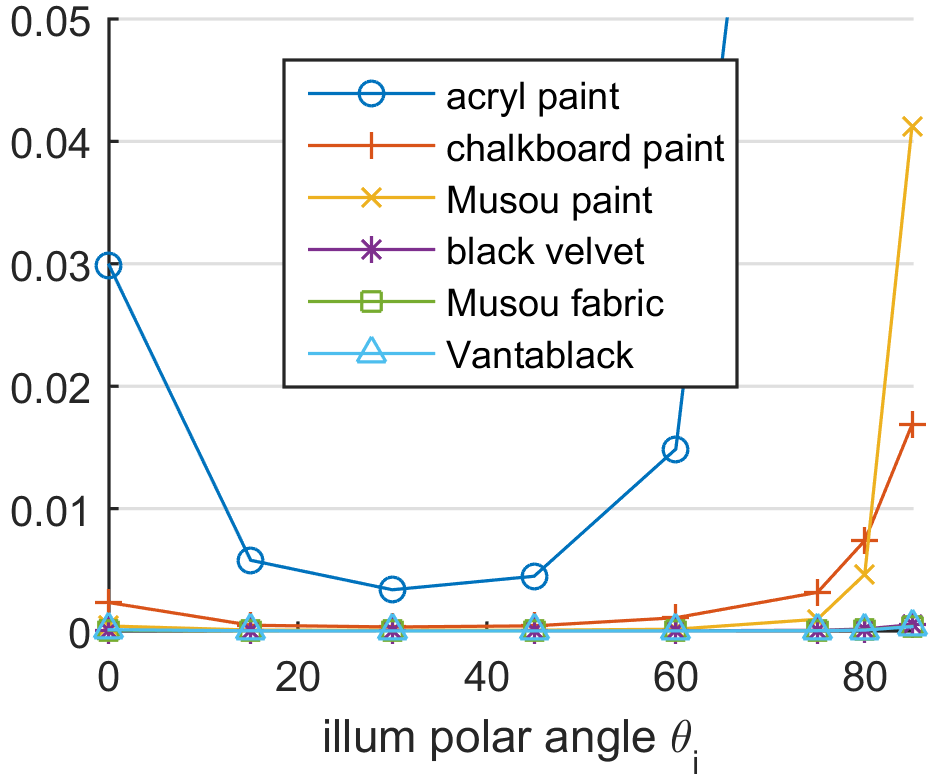} \\
\end{tabular}
\end{center}
\vspace{-6mm}
\caption{\label{fig:tis}Total integrated scattering, total hemispherical and specular reflectance.}
\vspace{-3mm}
\end{figure}

Figure~\ref{fig:tis} compares total integrated scatter (TIS), total hemispherical reflectance (THR), and the specular ($R_s$) reflectance component as functions of the illumination elevation angle $\theta_i$. Among the tested materials, the lowest TIS values are consistently observed for \emph{black velvet}, followed closely by \emph{Vantablack} and \emph{Musou fabric}, indicating strong suppression of directional scattering across a wide range of illumination angles. In terms of total hemispherical reflectance, \emph{Vantablack} and \emph{Musou fabric} achieve the lowest THR values over most illumination angles, with \emph{black velvet} exhibiting slightly higher reflectance due to its elevated diffuse component.

A notable trend is the increase in THR for \emph{acrylic paint} at low and moderate illumination polar angles, which is consistent with its strong specular contribution and reduced light trapping. In particular, coating-based materials show significantly higher $R_s$ values that increase rapidly toward grazing illumination, whereas fabric-based and ultra-black materials maintain low $R_s$ across all angles, due to their fibrous microstructure enhancing multiple scattering.

\subsection{Visual appearance in rendered scenes}

\begin{figure}[!ht]
\begin{center}
\renewcommand{\arraystretch}{0.4} 
\begin{tabular}{p{0.1cm}p{3.7cm}p{3.7cm}p{3.7cm}}
& \small{nominal light intensity} & \small{nominal scaled 10$\times$} & \small{nominal scaled 100$\times$} \\
\rotatebox{90}{\small{acryl paint}} &
\includegraphics[width=0.3\columnwidth]{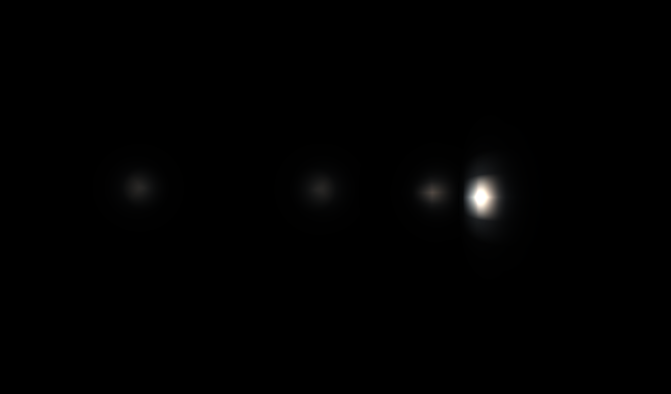} &
\includegraphics[width=0.3\columnwidth]{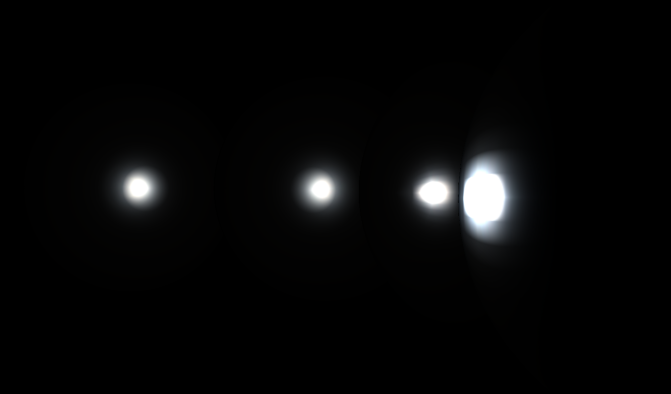} &
\includegraphics[width=0.3\columnwidth]{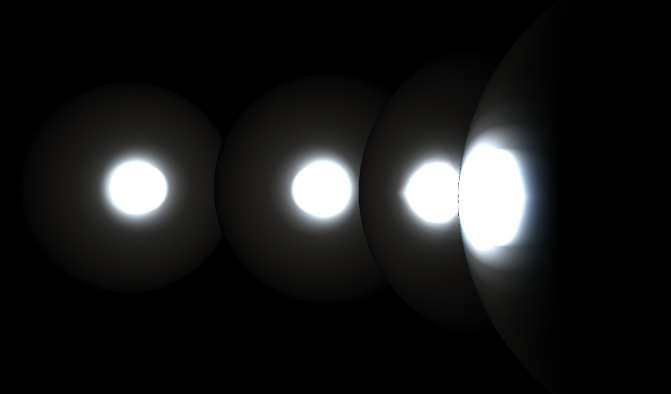} \\
\rotatebox{90}{\small{chalkboard paint}} &
\includegraphics[width=0.3\columnwidth]{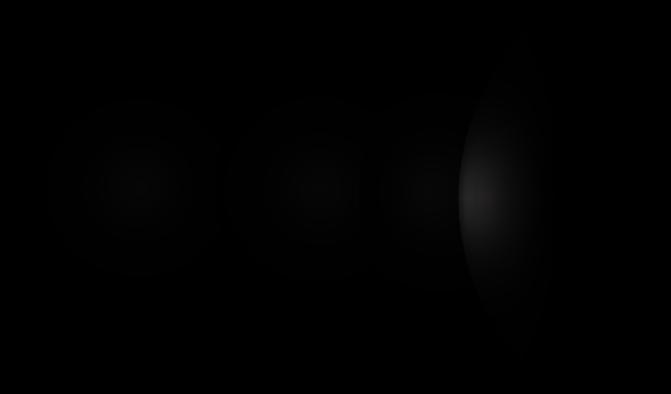} & 
\includegraphics[width=0.3\columnwidth]{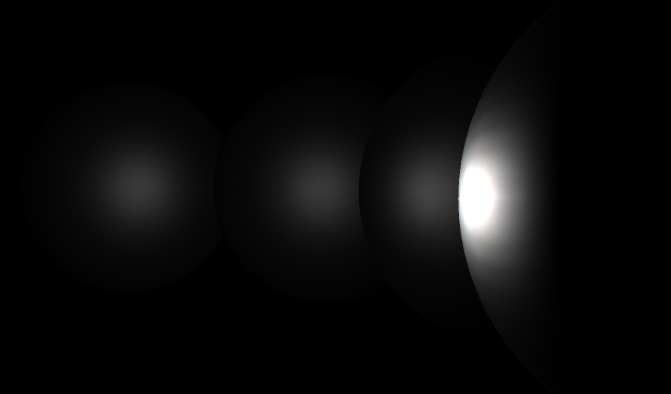} & 
\includegraphics[width=0.3\columnwidth]{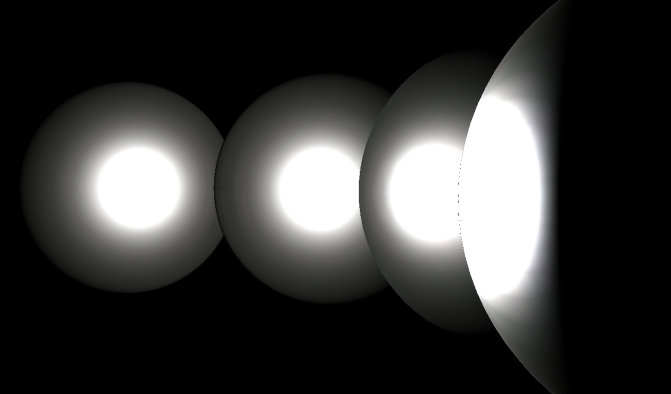} \\
\rotatebox{90}{\small{Musou paint}} &
\includegraphics[width=0.3\columnwidth]{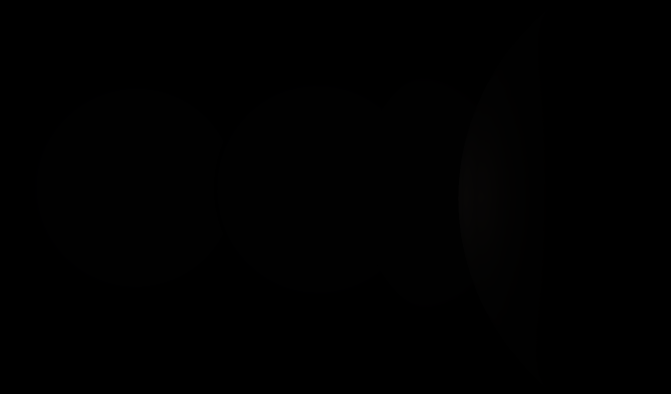} & 
\includegraphics[width=0.3\columnwidth]{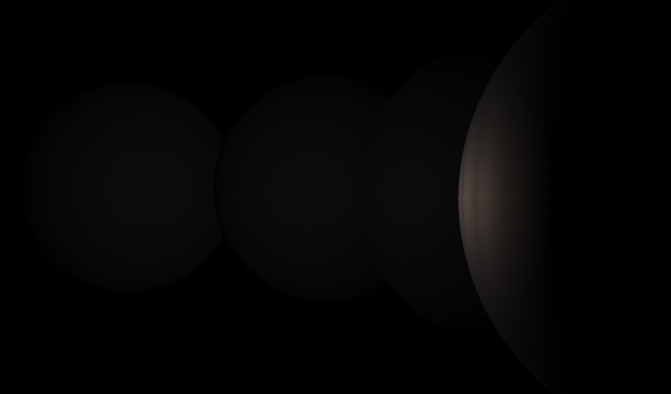} &
\includegraphics[width=0.3\columnwidth]{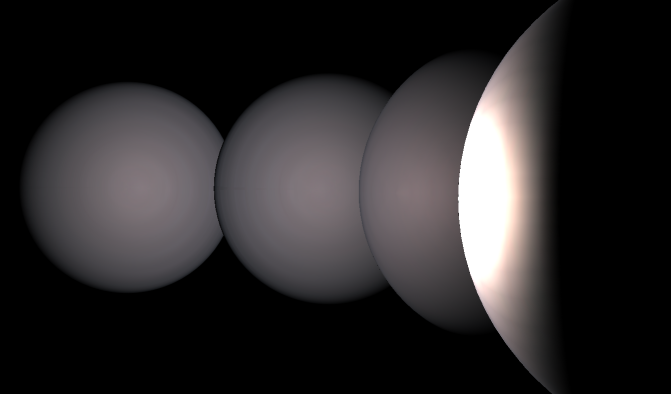} \\
\rotatebox{90}{\small{black velvet}} &
\includegraphics[width=0.3\columnwidth]{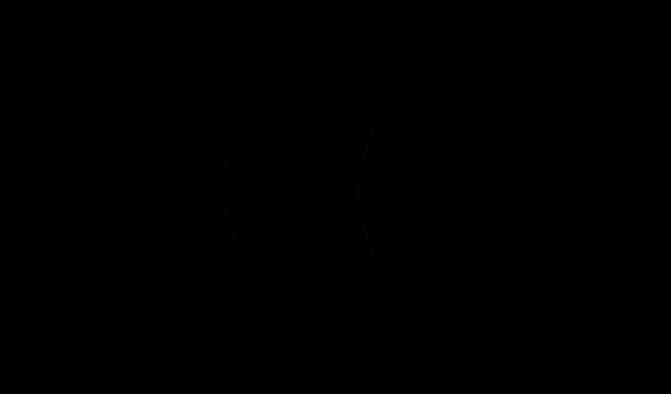} &
\includegraphics[width=0.3\columnwidth]{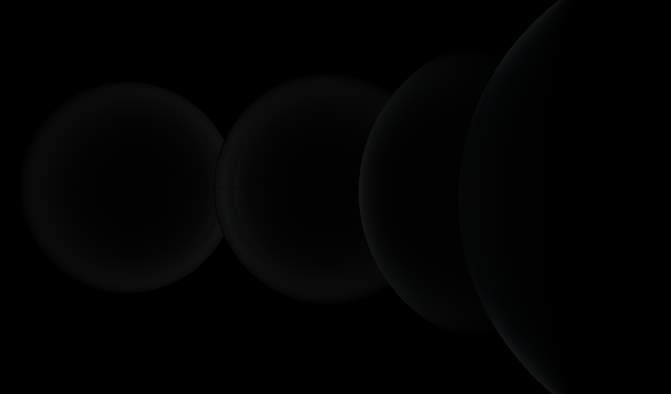} &
\includegraphics[width=0.3\columnwidth]{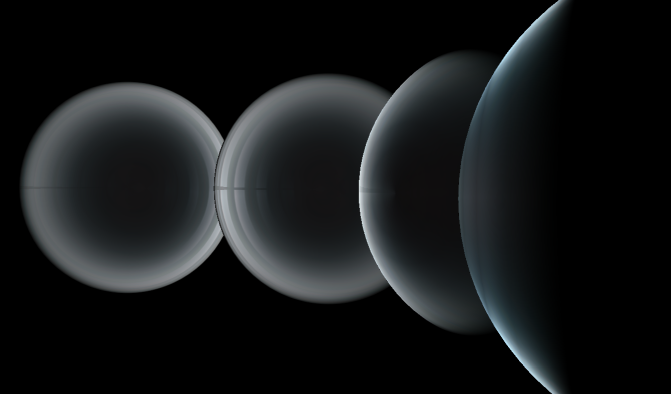} \\
\rotatebox{90}{\small{Musou fabric}} &
\includegraphics[width=0.3\columnwidth]{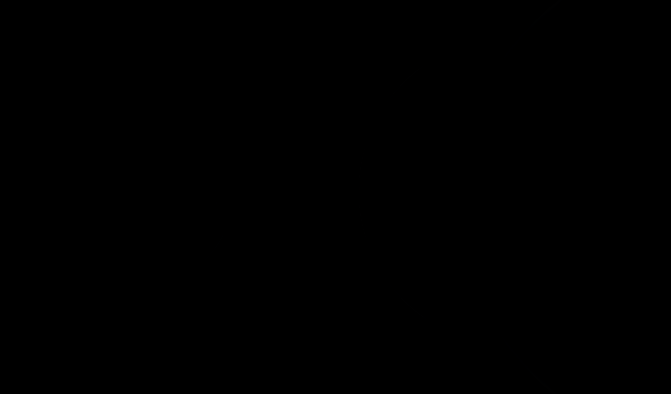} &
\includegraphics[width=0.3\columnwidth]{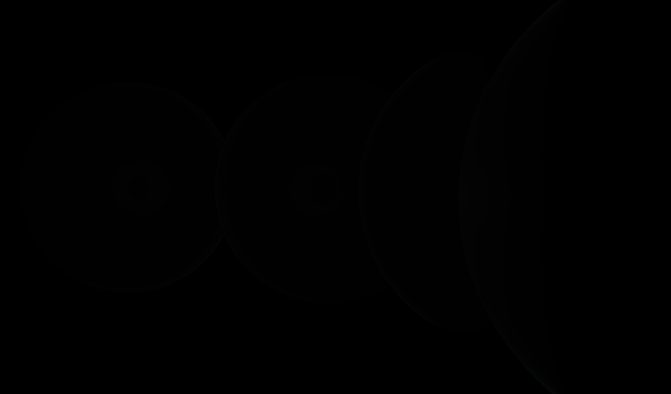} &
\includegraphics[width=0.3\columnwidth]{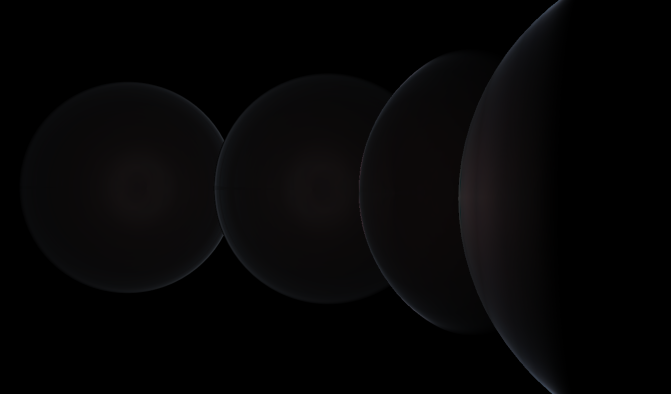} \\
\rotatebox{90}{\small{Vantablack}} &
\includegraphics[width=0.3\columnwidth]{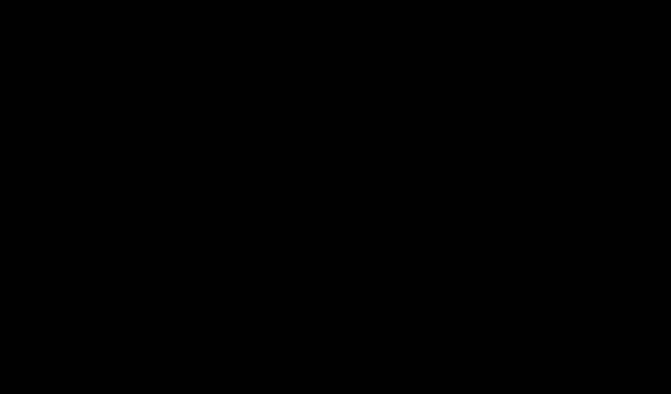} &
\includegraphics[width=0.3\columnwidth]{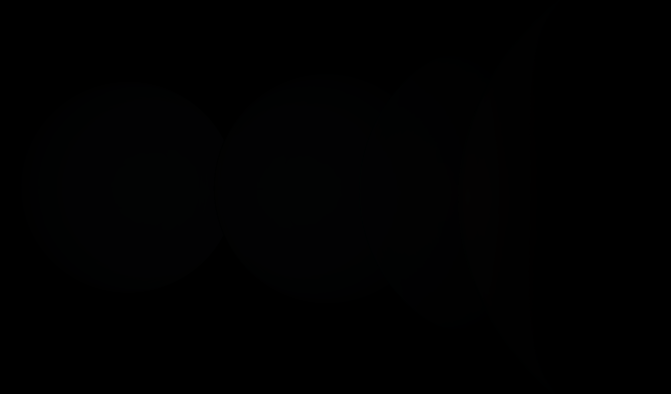} &
\includegraphics[width=0.3\columnwidth]{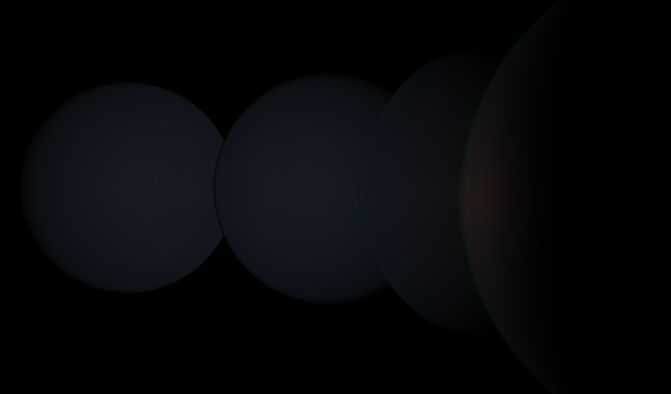} \\
\end{tabular}
\end{center}
\vspace{-5mm}
\caption{\label{fig:render}Renderings of the test scene consisting of four spheres illuminated by a point light source at nominal intensity and at increased intensities scaled by factors of 10 and 100.}
\vspace{-3mm}
\end{figure}

Fig.~\ref{fig:render} shows renderings of the test scene at nominal intensity (compare to 99\% reflectance in Fig.~\ref{fig:scene}(b)) and at increased intensities scaled by factors of 10 and 100. 

We first compare the coated samples. Among these, \emph{acrylic paint} exhibits the strongest specular response, which is already visible at nominal illumination and becomes increasingly pronounced at higher light intensities. The non-circular appearance of the highlights suggests that the observed reflection may not be solely determined by the paint layer itself, but could be influenced by local variations in coating thickness or coverage, potentially allowing partial reflection from the underlying polished aluminum substrate. In contrast, \emph{chalkboard paint} shows a noticeably reduced specular intensity accompanied by a broader angular distribution, which is consistent with increased surface roughness leading to more diffuse redistribution of reflected energy. \emph{Musou paint} exhibits comparatively low reflectance over most illumination and viewing directions, with a gradual increase in reflected intensity primarily observed at grazing angles, indicating a more angularly uniform response among the tested coatings.

We next consider the fabric-based materials. The reference \emph{black velvet} demonstrates effective suppression of mirror-like specular reflection relative to all coatings; however, it exhibits a visible sheen that becomes more apparent at higher illumination levels. This behavior is characteristic of fibrous materials and is commonly attributed to grazing-angle scattering from aligned surface fibers, which is particularly noticeable near the silhouettes of the spheres. A similar effect is observed for \emph{Musou fabric}, but with substantially reduced intensity, resulting in a more uniform appearance across viewing directions. In this case, the residual reflectance remains comparable to or lower than the specular component observed for \emph{Musou paint}, suggesting improved light trapping within the fabric structure.

Finally, the reference \emph{Vantablack} sample exhibits only faintly visible reflectance even at illumination levels increased by a factor of 100, indicating a consistently low reflected intensity across most angles. While the overall magnitude of its reflectance appears comparable to that of \emph{Musou fabric} in this visualization, \emph{Vantablack} shows the most spatially uniform angular response among all tested materials. Minor deviations from this uniformity are observable only at very low grazing angles, which is consistent with previously reported behavior of VACNT surfaces \cite{Lehman2017}.

\subsection{Psychophysical evaluation}
\begin{figure}[!ht]
\begin{center}
\includegraphics[width=0.75\columnwidth]{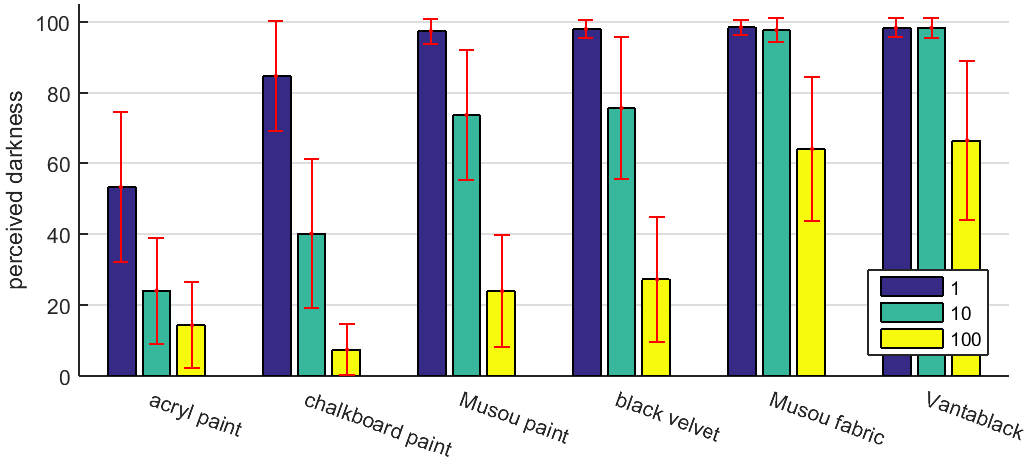} \\
\end{center}
\vspace{-5mm}
\caption{\label{fig:psycho}Perceived darkness of the tested materials as a function of lighting intensity scale (color-coded bars). Error bars indicate the standard deviation across observers.}
\vspace{-3mm}
\end{figure}

Finally, we used the rendered scenes shown in Fig.~\ref{fig:render} to assess human perception of material darkness. A total of 36 participants rated the perceived darkness of each material using a slider ranging from 0 (most reflective/bright) to 100 (most ultra-black). To help participants anchor the scale, examples of extremely bright and extremely dark materials were shown prior to the experiment. Each participant completed 18 trials corresponding to the 6 materials evaluated under the three lighting intensities, as shown in Fig.~\ref{fig:render}. The mean study duration was approximately 5 minutes. Inter-observer consistency was high: correlations between individual participants and the median observer ranged from 0.726 to 0.983, indicating robust agreement across observers.
The obtained psychophysical ratings shown in Fig.~\ref{fig:psycho} reveal clear and consistent perceptual differences among the tested black materials, strongly modulated by lighting intensity. Under the lowest illumination level (intensity 1), observers rated all materials as relatively dark, with ultra-black surfaces such as \emph{Musou fabric} and \emph{Vantablack} achieving the highest perceived darkness. As the intensity increases, perceived darkness decreases across all materials, but at markedly different rates: standard coatings like \emph{acrylic paint} and \emph{chalkboard paint} show large drops in perceived darkness, whereas \emph{black velvet}, \emph{Musou paint}, \emph{Musou fabric}, and \emph{Vantablack} maintain substantially higher darkness ratings even at intensities 10 and 100. Notably, \emph{Vantablack} and \emph{Musou fabric} remain perceptually darkest across all intensities, confirming their superior ability to suppress visible reflections under challenging lighting conditions. These findings align with the measured BRDF and TIS properties, demonstrating a strong correspondence between physical reflectance characteristics and human perceptual judgments.

\section{Discussion and limitations}

The presented analysis combines dense BRDF measurements, physically based rendering, and psychophysical evaluation to characterize the optical behavior of black materials across a wide range of illumination and viewing geometries. Several limitations should be noted.
First, the BRDF measurements were acquired under controlled laboratory conditions and for a fixed wavelength range in the visible spectrum. Although this is appropriate for most imaging and calibration applications, material behavior may differ outside the measured spectral range, particularly in the near-infrared, where absorption mechanisms and subsurface scattering can change. Second, the rendering experiments rely on a simplified geometric scene and a single point light source to emphasize angular scattering effects. While this configuration is well suited for comparative analysis, more complex illumination environments may lead to different perceptual responses.
The psychophysical evaluation was conducted using rendered images shown on LDR screen rather than physical samples, which allows precise control of illumination and geometry but may not capture all aspects of real-world viewing conditions, such as surface wear, contamination, or large-scale spatial context. In addition, the perceptual task focused on perceived darkness, which represents a dominant but not exclusive aspect of material appearance. Other perceptual attributes, such as texture, gloss impression, or visual comfort, were not explicitly assessed.

Despite these limitations, the consistency observed between BRDF-derived metrics, rendered appearance, and human perceptual judgments suggests that the reported trends are robust and relevant for practical material selection.

\section{Conclusions}
We presented a comparative analysis of black materials based on dense BRDF measurements, total scattering metrics, physically based rendering, and psychophysical evaluation to enable a coherent interpretation of material behavior across physical reflectance properties, visual appearance, and human perception.
Our results demonstrate that different classes of black materials suppress reflected light through distinct mechanisms: ultra-black and fabric-based materials consistently attenuate both diffuse and specular reflection across illumination and viewing angles, whereas conventional dark coatings exhibit pronounced angle-dependent increases in specular reflectance that dominate their hemispherical response at grazing illumination. These physical differences are reflected in rendered appearance and are corroborated by psychophysical ratings of perceived darkness.
The agreement between BRDF-derived measures, visual renderings, and observer judgments allows the results to be translated into practical guidance for material selection. Ultra-black materials provide the highest level of light attenuation but are often constrained by cost and mechanical fragility. Dark fabrics offer a robust and effective alternative in applications where space and environmental conditions permit, achieving high perceived darkness while maintaining low angular reflectance. Conventional dark coatings may be suitable in scenarios where illumination and viewing geometries avoid near-specular configurations, but their performance is inherently limited by stronger angle-dependent scattering.
Future work will extend this analysis to polarization-resolved measurements and spectral ranges beyond the visible domain.

\section*{Funding}
This research was supported by the Czech Science Foundation under grant GA22-17529S.

\section*{Disclosures}
The authors declare no conflicts of interest.

\section*{Data availability}
BRDF measurements and rendering code are available as a supplementary material.

\bibliography{references}

\end{document}